\documentclass[epj,nopacs]{svjour}

\usepackage{epsfig}

\newcommand{\be}{\begin{eqnarray}}
\newcommand{\ee}{\end{eqnarray}}

\begin{document}
\title{Saturation physics and angular correlations at RHIC and LHC }
\author{Jamal Jalilian-Marian} 
\institute{Department of Natural Sciences, Baruch College, 17 Lexington Ave., 
New York NY 10010 \hfill BCCUNY-HEP-08-02}
\date{}
%
\abstract{
We investigate the angular correlation between pions and photons produced in 
deuteron-gold collisions at RHIC and proton-lead collisions at LHC using the 
Color Glass Condensate formalism and make predictions for the dependence of 
the production cross section on the angle between the pion and photon at 
different rapidities and transverse momenta. Measuring this dependence would 
shed further light on the role of high gluon density and saturation dynamics 
at RHIC and LHC.
} 
\maketitle
\section{Introduction}
\label{intro}
The Color Glass Condensate (CGC) formalism has been quite successful in describing single inclusive 
production as well as multiplicity of hadrons at RHIC \cite{review}. While CGC based predictions 
for the suppression of forward rapidity spectrum of produced hadrons in deuteron-gold collisions 
were observed in the experimental data \cite{supp}, there have been more recent and more traditional approaches which can now also fit the data \cite{other}. Therefore, it is desirable to investigate more processes which could help verify/falsify the claim that saturation dynamics has been conclusively discovered at RHIC . Electromagnetic processes \cite{em} are specially promising tools in investigating saturation physics since they do not suffer from the theoretical uncertainties associated with hadronization. Another venue which is particularly sensitive to saturation effects, and
can serve a powerful discriminator between different approaches is two particle 
correlations \cite{twohadcorr}.

Two particle correlations are expected to be a sensitive probe of high energy QCD and saturation dynamics where one expects a decorrelation between the two hadrons when the rapidity separation between the two hadrons increases. This is due to the large phase space (in $x$ or rapidity) between the produced hadrons. Since the small $x$ evolution involves a random  walk in transverse momentum, this washes away the angular correlation which is the opposite of the case in pQCD in the DGLAP limit where the two produced hadrons are back to back. 

While two hadron production cross section has been investigated using the CGC formalism and analytic results are known \cite{twohad}, the resulting expressions are complicated, specially when the two hadrons are separated by large rapidities. In order to get results which lend themselves easily to numerical methods, one needs to make further assumptions and approximations which reduce the accuracy of the results and diminish the predictability of the formalism. For instance, the two hadron production cross section 
involves higher point functions of Wilson lines unlike the single inclusive production which involves only the two point function. One would then need to either solve the complicated JIMWLK \cite{review} equations for these higher point functions or make a mean field (or Gaussian) approximation where all higher point functions are approximated as powers of the two point function. This would clearly neglect some correlations which may be important. 

Here we investigate photon-pion production in deuteron-gold (or proton-lead) collisions which is a much simpler problem than two hadron production but still has the essential physics of saturation. The basic building block in this process is the quark anti-quark dipole scattering probability, denoted $N_F$, which also shows up in single inclusive photon or dilepton production, as well as single inclusive hadron production in the forward rapidity region. The dipole scattering probability satisfies the JIMWLK (or BK in large $N_c$ and mean field approximation) and its properties are understood quite well even. Even though an analytic solution to the dipole evolution equations is not known, it can be solved 
numerically \cite{stasto} and approximate analytic solutions are known in various limit.

\section{The production cross section}
\label{cs}
The cross section for inclusive production of a photon and a parton in the scattering of a 
quark on a given target $T$ such as a proton or nucleus is given by \cite{em}
\be
{d\sigma^{q\,T\rightarrow q\, \gamma\, X}\over d z\, d^2 l_t\, d^2 k_t\, d^2 b_t} &= &{e^2\over (2\pi)^5}
{z [1 + (1 - z)^2] \over [z \vec{l}_t - (1-z) \vec{k}_t]^2}\, {(\vec{l}_t + \vec{k}_t)^2 
\over k_t^2} \nonumber \\   
&& N_F (x_g, b_t, |\vec{l}_t + \vec{k}_t|)
\label{eq:cs_parton}
\ee
where $l_t$, $k_t$ are the transverse momenta of the final state quark and photon respectively, 
$z$ is the fraction of the energy of the initial state quark carried by the photon and $x_g$ is
the momentum fraction of the gluons in the target $T$ (proton or nucleus) given by 
\be
x_g = {1\over \sqrt{s}}\, [l_t\, e^{-y_q} + k_t\, e^{-y_{\gamma}}]
\ee
where $y_q$ and $y_{\gamma}$
are the rapidities of the final state quark and photon. To relate (\ref{eq:cs_parton}) to the production cross section for a hadron and a photon in a deuteron (proton)-target scattering, we will convolute 
(\ref{eq:cs_parton}) with quark distribution functions of a proton (or deuteron, using isospin symmetry) given by CTEQ6 and pion fragmentation functions given by KKP and rewrite (\ref{eq:cs_parton}) as
(below we will consider impact parameter integrated cross sections only)  
\be
{d N^{p(d)\,T\rightarrow h\, \gamma\, X}\over d q_t^2\, d k_t^2\, d y_{\gamma}\, d y_h\, d \theta} &= &
a\, \int_{z_min}^1 \, {d z \over z^5} \, f_{q/p} (x_p, Q^2) \, D_{h/q} (z, Q^2) \nonumber \\
&& [z^2 + ({q^-\over q^- + z k^-})^2] \, 
{(\vec{q}_t + z \vec{k}_t)^2\over (k^- \vec{q}_t - q^- \vec{k}_t)^2}\nonumber \\  
&& N_F (x_g, |\vec{q}_t/z + \vec{k}_t|)
\ee
with $q_t = z\, l_t$ and where 
$a = {e_q^2\, \alpha_{em}\over 2\, \sqrt{2}\, (2\pi)^3}\,{q^-\, e^{2 y_{\gamma}} \over \sqrt{s}}$ 
is a angle independent constant which will drop out when we consider the angular correlations below and $\vec{q}_t$, $q^-$ are the transverse momentum and energy of the produced pion. The angle $\theta$ is defined to be the angle between the produced photon and hadron. The lower limit of the $z$ integration is given by 
$z_{min} = {q_t\over \sqrt{s}}\,e^{y_h}\, [1 - {k_t\over \sqrt{s}}\, e^{y_{\gamma}}]^{-1}$ and 
$x_p = {1\over \sqrt{s}}\, (q_t/z \, e^{-y_h} + k_t\, e^{-y_{\gamma}})$. Transverse momenta and energies and energies of the produced pion are related via $q^- = {q_t\over \sqrt{s}}\, e^{y_h}$
and similarly for the photon.

Before a quantitative numerical analysis it is worth looking at some limits of (\ref{eq:cs_parton}). It can be shown \cite{em} that it the limit where the photon has a very large transverse momentum (such that the 
collinear photon-quark contributions are suppressed), one recovers the standard pQCD result for 
$q\, g \rightarrow q\,\gamma$ convoluted with the gluon distribution function of the target. On the 
other hand and in the collinear limit, one recovers the fragmentation photons \cite{jjm}. Therefore, the above expressions include contributions of both direct and fragmentation photons. On the other hand, it does not include the contribution of gluons from the projectile scattering from the target and then fragmenting into photons which necessarily involves quark production in the intermediate stage and is therefore higher order in $\alpha_s$ even though it may be numerically important in mid rapidity region. For the sake of theoretical consistency, this contribution is not included here.

Therefore, for the case when the produced pion and photon have the same transverse momentum and rapidity, one expects a collinear divergence as the angle $\theta \rightarrow 0$ in the correlation function which disappears as one goes to larger angles where in the limit where $\theta \rightarrow 180$ one recovers the standard back to back configuration. If the pion and photon are separated by a large rapidity while at the same transverse momentum, the cross section does not receive any contribution from collinear configurations.   

\subsection{Numerical results}
\label{numerics}
In order to investigate the angular correlation between the produced pion and photon, we define the 
correlation function $C (\theta)$ as
\be
C (\theta) \equiv {
{d N^{p(d)\,T\rightarrow h\, \gamma\, X}\over d q_t^2\, d k_t^2\, d y_{\gamma}\, d y_h\, d \theta}
\over
\int d \theta {d N^{p(d)\,T\rightarrow h\, \gamma\, X}\over d q_t^2\, d k_t^2\, d y_{\gamma}\, d y_h\, d \theta}
}.
\ee
We expect that $C (\theta ) \rightarrow 1$ when $\theta \rightarrow 0$ when the pion and photon have the same rapidity and transverse momentum since the cross section in this case is dominated by the collinear divergence between the photon and pion.

\begin{figure}[htbp]
\begin{center}
\includegraphics[width=8.8cm]{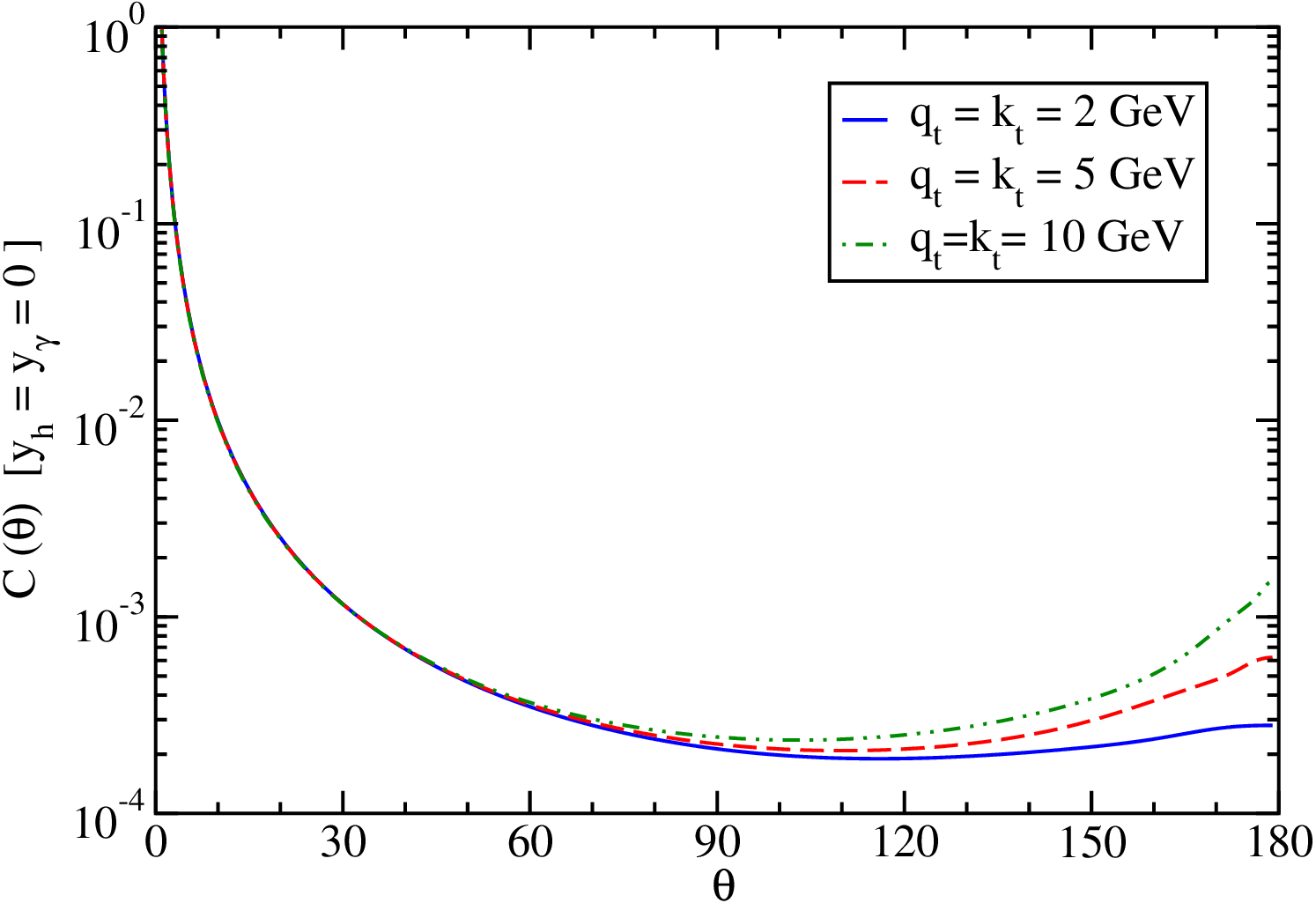}
\caption{Pion-photon correlation function in deuteron-gold collisions at RHIC.}
\protect\label{fig:dA_yhyg0}       
\end{center}
\end{figure}

In Fig. (\ref{fig:dA_yhyg0}) we show the correlation function $C (\theta )$ when both photon and pion are in mid rapidity and for several transverse momenta, for deuteron-gold collisions at RHIC. It goes to $1$ when the angle $\theta$ goes to $0$ while at $\theta \rightarrow 180$, its magnitude increases as one goes to higher transverse momentum. We note that the magnitude of the correlation function changes by almost an order of magnitude as one increases the transverse momenta of the pion and photon from 
$2\, $ GeV to $10\, $ GeV when $\theta \rightarrow 180$ where the $q\, g \rightarrow q\, \gamma$ process dominates. This increase of the correlation function is due to the fact that as one goes to higher transverse momenta, high gluon density effects become weaker.

\vspace{0.2in}

\begin{figure}[htbp]
\begin{center}
\includegraphics[width=8.8cm]{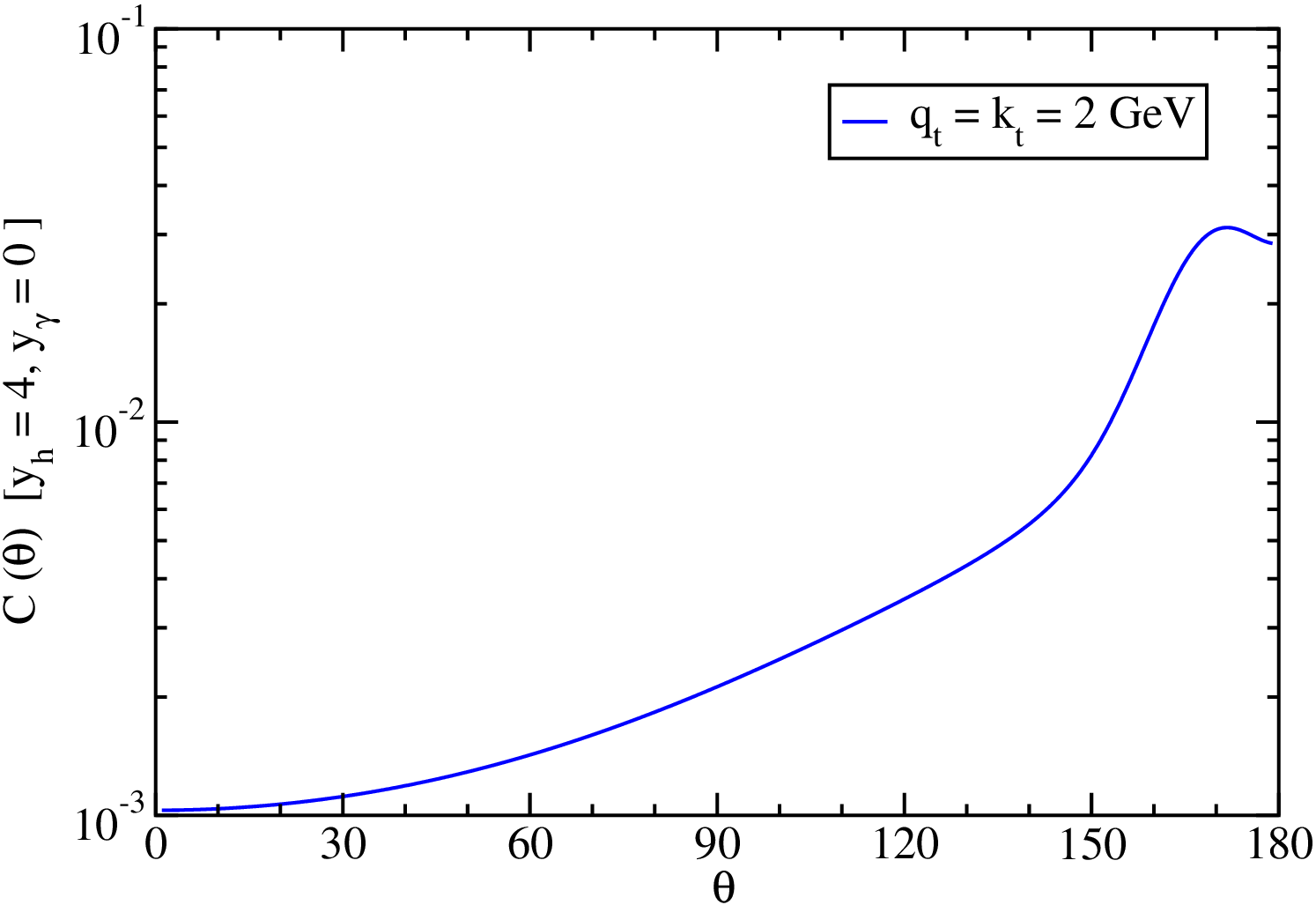}
   \caption{Same as in Fig. (\ref{fig:dA_yhyg0}) but with different rapidities.}
   \protect\label{fig:dA_yh4yg0}
\end{center}
\end{figure}

\begin{figure}[bhtp]
\begin{center}
\includegraphics[width=8.8cm]{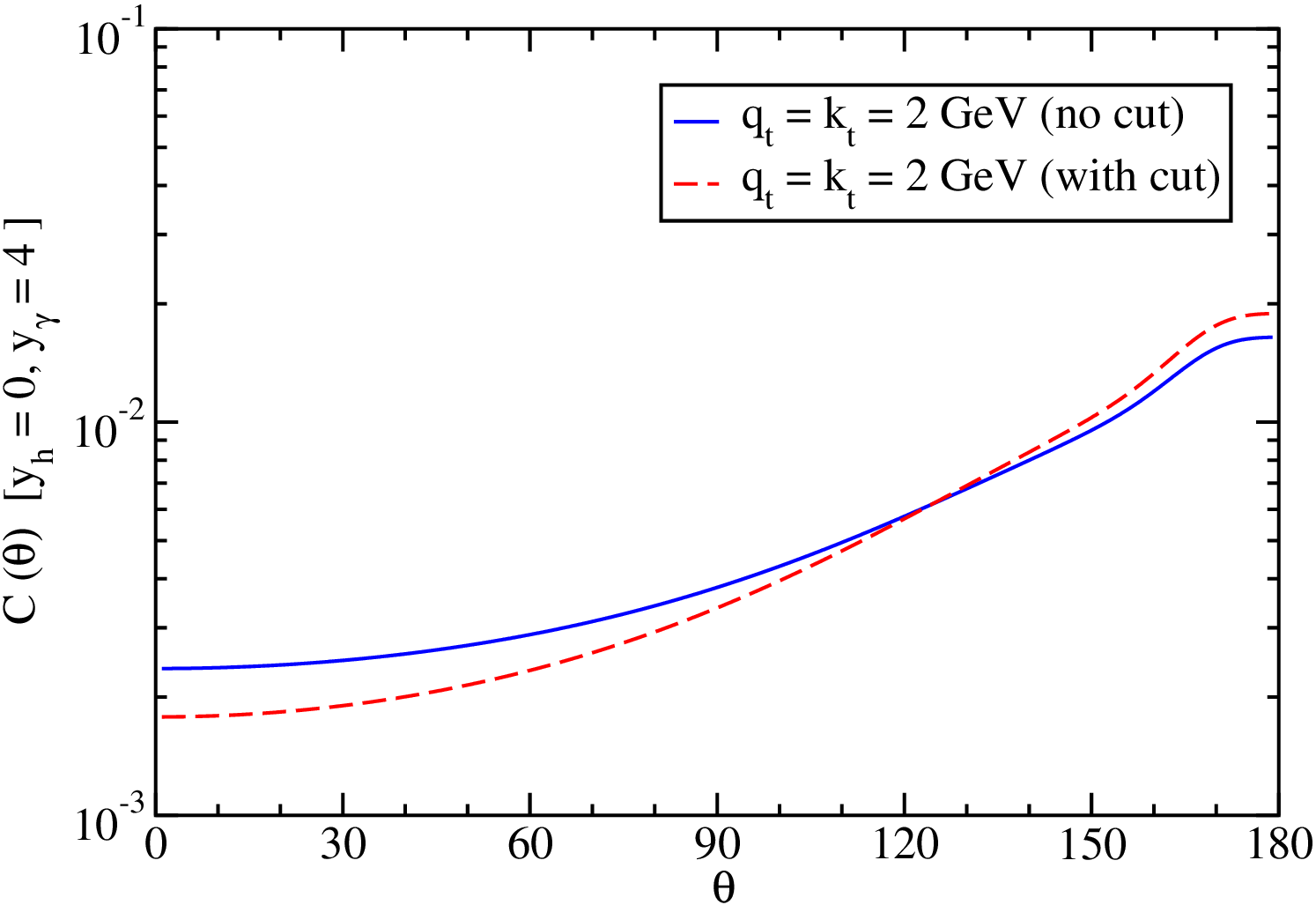}
\caption{Same as in Fig. (\ref{fig:dA_yhyg0}) but with different rapidities 
and with or without an isolation cut.}
\protect\label{fig:dA_yh0yg4}       
\end{center}
\end{figure}

In Fig. (\ref{fig:dA_yh4yg0}) we show the correlation function $C (\theta )$
in the case when the photon and pion are separated by a large rapidity ($y_h = 4$, $y_\gamma = 0$) but are at the same transverse momenta. As discussed earlier, the contribution of fragmentation photons becomes negligible and the 
correlation function is dominated by the quark-gluon scattering. Unfortunately, the transverse momentum
reach in the forward rapidity region of RHIC is limited to low transverse momenta due to the kinematics.
We also note that the small dip in the region when $\theta \rightarrow 180$ is numerical artifact and is caused by the rapid oscillation of the dipole profile which makes it extremely difficult to Fourier transform it accurately \footnote{We than F. Gelis for the use of his program for Fourier transforms.}.
In Fig. (\ref{fig:dA_yh0yg4}) we show the correlation function when the two particles are separated by a large rapidity but with $y_h = 0$ and $y_\gamma = 4$. We also show the effect of an isolation cut on the photon where the cut parameter $\epsilon$ is independent of the cone size at Leading Order and is taken to be $\epsilon = 0.1$ at RHIC. This cut affects the minimum of the $z$ integration in the cross section
to 
\be
z_{min}^{cut} = {1\over \epsilon} {q_t\over k_t}\, e^{y_h - y_{\gamma}}
\ee
It is clear that the isolation cut would severely diminish the cross section unless the rapidity
separation between the pion and photon is large and negative, or when the transverse momenta of the
two particles differ by orders of magnitude. Therefore, the isolation cut is not applied to the results
shown in the other figures.

\vspace{0.2in}

\begin{figure}[hbtp]
\begin{center}
\includegraphics[width=8.8cm]{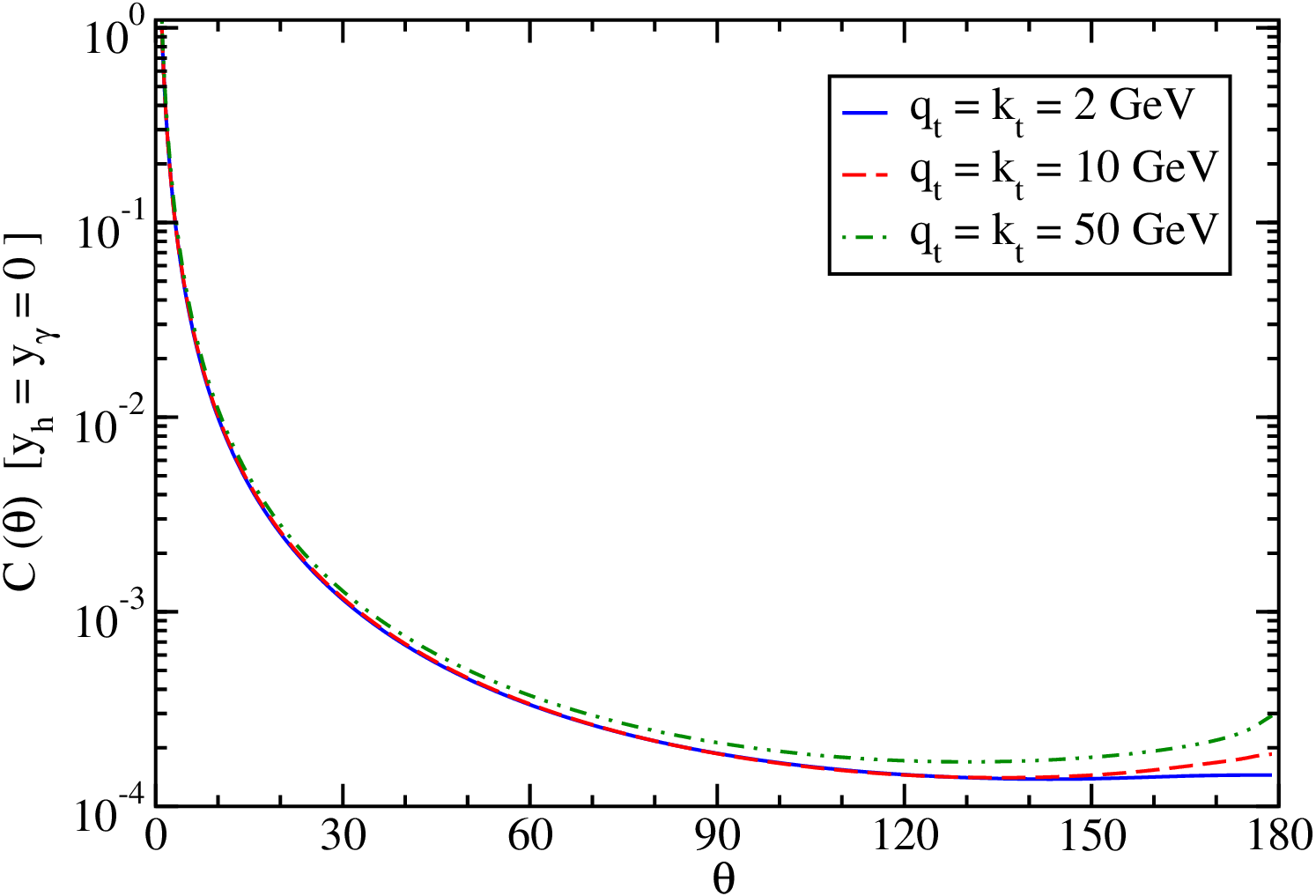}
\caption{Pion-photon correlation function in mid rapidity proton-lead collisions at LHC.}
\protect\label{fig:pA_yhyg0}       
\end{center}
\end{figure}

Finally in Fig. (\ref{fig:pA_yhyg0}) we show the correlation function for proton-lead collisions at 
LHC energies ($\sqrt{s} = 8800\, $ GeV). A similar pattern in the angle dependence is observed except that the transverse momentum dependence of the correlation function is much weaker in the back to back case. This is presumably due to the larger kinematic region in LHC where leading twist shadowing is more  important as compared to RHIC. In summary, hadron-photon angular correlation function provides an additional tool with which to investigate saturation dynamics at RHIC and LHC. It depends on the same basic ingredient as photon and dilepton production and single hadron production in the forward rapidity region. Therefore an experimental verification/falsification of this correlation would shed much light on the domain of applicability of CGC and saturation physics.

\vspace{0.2in}

\end{document}